\documentclass[11pt]{article}

\usepackage{amssymb,amsmath,amsthm}

\theoremstyle{plain}
\newtheorem{prop}{Proposition}[section]
\newtheorem{thm}[prop]{Theorem}
\newtheorem{cor}[prop]{Corollary}
\newtheorem{lem}[prop]{Lemma}

\theoremstyle{remark}

\newcommand{\ds}{\displaystyle}

\newcommand{\wt}{\widetilde}

\begin{document}

\title{Grover's Algorithm for Multiobject Search in Quantum Computing}

\author{Goong Chen$^{1,2}$, Stephen A.~Fulling$^{1}$, 
and Marlan  O.~Scully$^{3,4}$}

\date{}
\maketitle

\abstract{L. K.~Grover's search algorithm in quantum computing gives an
optimal,  square-root speedup in the search for a single object in a large
unsorted  database. In this paper, we
expound  Grover's algorithm in a Hilbert-space 
 framework that isolates its geometrical essence, 
 and we generalize it to the case where  more than one object
satisfies the search criterion.}

 \vskip1truein

 \centerline{\it In memory of Daniel Walls}
\vfil

\begin{itemize}
\item[1.] Department of Mathematics, Texas A\&M University, College Station, TX 
\ 77843-3368.
\item[2.] Supported in part by Texas A\&M University Interdisciplinary Research 
Grant IRI 99-22.
\item[3.] Department of Physics
and Institute for Quantum Studies, Texas A\&M University, College Station, TX \ 
77843-4242, and Max-Planck Institut f\"ur Quantenoptik, 
Munich.
\item[4.] Supported in part by  ONR, NSF, and Welch Foundation.
\end{itemize}

\eject

\baselineskip = 18pt

\section{Introduction}\label{grov:sec1}

\indent

A quantum computer (QC) is envisaged as a collection of 2-state ``quantum 
bits''\negthinspace,
  or {\em qubits\/} (e.g., spin 1/2 particles). 
 Quantum computation does 
calculations on data densely coded in the 
 entangled states that are the hallmark of 
quantum mechanics, potentially yielding unprecedented parallelism in 
computation, as P.~Shor's work on factorization 
[\ref{shor1},~\ref{shor2}] proved in 1994. 
Two years later, L.~K.~Grover [\ref{grover1}]  showed that for an 
unsorted database with $N$ items in storage,
  it takes an average number of 
$\mathcal{O}(\sqrt N)$ searches to locate a single desired object 
by his quantum search algorithm. 
 If $N$ is a very large number,
 this is a significant square-root 
speedup over the exhaustive search algorithm in a classical 
computer, which requires an average number of $\frac{N+1}2$ 
searches. Even though Grover's algorithm is not logarithmically 
fast (as Shor's is), it has been argued that
 the wide range of its applicability compensates for this 
[\ref{BHT}]. 
Furthermore, the quantum speedup of the search algorithm
is {\em indisputable},
whereas for factoring the nonexistence of competitively
fast classical algorithms has not yet been proved
[\ref{BL1},~\ref{BL2}].

Grover's original papers [\ref{grover1}, \ref{grover2}] deal with 
 search for a single object. 
 In practical applications, typically more than one item  will 
satisfy the criterion used for searching.
In  the simplest generalization of Grover's algorithm,
 the number of ``good'' items is known in advance 
 (and greater than~$1$).
 Here we expound this generalization, 
along the lines of a treatment of the single-object 
 case by Farhi and 
Gutmann [\ref{FG}, Appendix] that makes the 
 Hilbert-space geometry of the situation very clear.

 The success of Grover's algorithm and its multiobject generalization 
 is attributable to two main sources:
\begin{itemize}
\item[(i)] the notion of amplitude amplication; and
\item[(ii)] the dramatic reduction to invariant subspaces of low dimension for 
the unitary operators involved.
\end{itemize}
 Indeed, the second of these can be said to be responsible for the 
first:
 A proper geometrical formulation of the process shows that all the 
``action'' takes place within a {\em two-dimensional, real\/} 
subspace of the Hilbert space of quantum states.
 Since the state vectors are normalized, the state is confined to a 
one-dimensional unit circle and (if moved at all) initially has 
nowhere to go except toward the place where the amplitude for the 
sought-for state is maximized.
 This accounts for the robustness of Grover's algorithm --- that 
is,
 the fact that Grover's original choice of initial state and of the
 Walsh--Hadamard transformation can be replaced by (almost) any
 initial state and (almost) any unitary transformation
 [\ref{grover3}, \ref{jozsa},~\ref{BHT}].

The notion of amplitude amplification 
  was  emphasized  in the original works [\ref{grover1}, 
\ref{grover2}, \ref{grover3}] of Grover himself
 and in those of 
 Boyer, Brassard, H\o{}yer and Tapp [\ref{BBHT}] and 
 Brassard, H\o{}yer and Tapp~[\ref{BHT}].
(See also [\ref{BL1},~\ref{BL2}].)
 Dimensional reduction is prominent 
 in the papers by Farhi and Gutmann [\ref{FG}]  and Jozsa~[\ref{jozsa}]. 
We applied dimensional reduction to multiobject search
independently of references [\ref{BBHT}] and [\ref{BHT}] and later
learned that the same conclusions about multiobject search 
 (and more) had been 
obtained there in a different framework.
 (We modestly suggest that our framework is clearer.)

The rest of the paper is divided into two parts. In \S 2, we reformulate the 
original Grover algorithm, and in \S 3, a multiobject search algorithm is 
studied.

\section{Introduction to Grover's Algorithm}\label{grov:sec2}

\setcounter{equation}{0}

\indent

In this section, we review Grover's algorithm for searching a 
single element in an unsorted database containing $N\gg1$ items, 
following  [\ref{FG}]. This proof is presented in a way that makes 
possible the generalization of the algorithm to perform multiobject 
search in an unstructured database.

Grover  treated the following abstract problem:
 We are given a Boolean function $f(a)$, $a=1$, $2$, $\ldots$,~$N$, 
which is known to be zero for all $a$ except at a single point,
  say at $a=w$, where $f(w)=1$.
 The problem is to find the  value $w$. 
 (The function is an ``oracle'' or ``black box'':
 all we know about it is its output for any input we care to 
insert.)
 On a 
classical computer we have to evaluate the function
 $\frac{N+1}2$ times on average 
to find the answer to this problem. In contrast, Grover's 
quantum algorithm finds $w$ in $\mathcal{O}(\sqrt{N})$ steps. 

The quantum-mechanical statement of the problem is that given an
 orthonormal basis 
$\left\{\,|\,a\rangle\,:\, a=1,2,\ldots,N\right\}$ 
 we want to single out the basis 
element $|\,w\rangle$ for which $f(w)=1$. 
 (More concretely,
 each $|\,a\rangle$ is to be an eigenstate of the qubits making up 
the~QC.  If $N=2^n$, then $n$ qubits will be needed.)
 At $t=0$, we prepare the state of the 
system $|\,\psi\rangle$ in a superposition of the states 
$\left\{\,|\,a\rangle\right\}$, each with the same probability:
\begin{equation}\label{grov:eq2.1}
|\,\psi\rangle = \frac{1}{\sqrt{N}}
  \sum_1^N |\,a\rangle \equiv |\,s\rangle .         
\end{equation}

 By the Gram--Schmidt construction we extend $|\,w\rangle$
 to an orthonormal basis for the subspace 
spanned by $|\,w\rangle$ and $|\,s\rangle$.
 That is, we introduce a normalized vector $|\,r\rangle$ orthogonal 
to $|\,w\rangle$, 
\begin{equation}\label{grov:eq2.2}
|\,r\rangle = \frac{1}{\sqrt{N-1}} \sum_{a\neq w} |\,a\rangle ,                
\end{equation}
 and find that
 the initial state has the representation
\begin{equation}\label{grov:eq2.3}
|\,s\rangle = \sqrt{\frac{N-1}{N}} |\,r\rangle +
  \frac{1}{\sqrt{N}} |\,w\rangle.        
\end{equation}

Following Grover, we now define the  unitary operator of
 {\em  inversion about average},
\begin{equation}\label{grov:eq2.4}
I_s = \pmb{I} - 2|s\rangle\langle s|.
\end{equation}
 Notice that the only action of this operator is to flip the sign 
of the state $|\,s\rangle$; that is, 
$I_s\,|\,s\rangle = - |\,s\rangle$ but
  $I_s|\,v\rangle = |\,v\rangle$ if $\langle s|v\rangle =0$. 
  Using (\ref{grov:eq2.3}) 
we write $I_s$ as
\begin{equation}\label{grov:eq2.5}
I_s = -\left(1-\frac{2}{N}\right)(\,|\,r\rangle\langle r\,|
  -|\,w\rangle\langle
w\,|\,) - 2  \frac{\sqrt{N-1}}{N}(\,|\,r\rangle\langle w\,| + 
|\,w\rangle\langle r\,|\,) .  
\end{equation}
 In other words, with respect to the orthonormal basis the operator 
$I_s$ is represented by the orthogonal (real unitary) matrix
 \begin{equation}\nonumber
\left[\begin{array}{cc}
1-\ds\frac2N&-2\cdot \ds\frac{\sqrt{N-1}}{N}\\   
\noalign{\smallskip}
-2\cdot \ds\frac{\sqrt{N-1}}{N}&-\left(1-\ds\frac2N\right)
 \end{array}\right].
 \end{equation}

Similarly, the  operator $I_w$ is defined by
\begin{equation}\label{grov:eq2.7} I_w = \pmb{I} - 
2\,|\,w\rangle\langle w\,|  	\end{equation} 
and satisfies    $I_w |\,w\rangle = -|\,w\rangle$.
 The crucial fact is that   in terms of the oracle function~$f$, 
\begin{equation}\label{grov:eq2.6}
I_w\,|\,a\,\rangle = (-1)^{f(a)}\,|\,a\rangle
\end{equation}
for each $|\,a\,\rangle$ in the original basis for the full state 
space of the~QC.
Therefore, to execute the operation $I_w$ one does not need to 
know~$w$; one only needs to know~$f$.
 (And conversely, being able to execute $I_w$ does not mean that 
one can immediately determine~$w$;
 $\sqrt{N}$ steps will be needed.)

A ``Grover iteration'' is the unitary operator $U \equiv -I_s I_w\,$.
This product can be calculated easily in either the bra-ket or the 
matrix formalism.
 In particular,
 for the transition element $\langle w\,|\,U\,|\,s\rangle$ we obtain
\begin{align}
\langle w\,|\,U\,|\,s\rangle& = \langle w\,|\,\left[\,
\left(1-\frac{2}{N}\right)\, \pmb{I}  + 
\frac{2\sqrt{N-1}}{N}(\,|\,w\rangle\langle r\,| -
|\,r\rangle \langle w\,|\,)\,\right]\, |\,s\rangle \nonumber \\
\label{grov:eq2.8}
       & =  \left(1-\frac{2}{N}\right)\frac{1}{\sqrt{N}} + 2 \left(1 
-\frac{1}{N}\right)\frac{1}{\sqrt{N}}\\ 
       &= \frac{1}{\sqrt{N}} + \frac{2}{\sqrt{N}} + 
\mathcal{O}(N^{-3/2}) . \nonumber
\end{align}

The fact that the matrix element $\langle w\,|\,U\,|\,s\rangle$ is 
nonzero can be used to reinforce the probability amplitude of the 
unknown state $|\,w\rangle$. If we use $U$ as our unitary search 
operation, then after $m\gg 1$ trials the value $\langle 
w\,|\,U^{m}\,|\,s\rangle$ can be evaluated as follows:
\begin{align*}
\langle w|U^m|s\rangle &= [1~~~0] \left[\begin{array}{cc}
1-\ds\frac2N&2\cdot \ds\frac{\sqrt{N-1}}{N}\\   
\noalign{\smallskip}
-2\cdot \ds\frac{\sqrt{N-1}}{N}&1-\ds\frac2N\end{array}\right]^m 
\left[\begin{array}{c}
\ds\frac1{\sqrt N}\\ 
\noalign{\smallskip}
\sqrt{\ds\frac{N-1}N}\end{array}\right]\\
&= [1~~~0] \left[\begin{array}{cr}
\cos\theta&\sin \theta\\ 
 - \sin\theta&\cos \theta\end{array}\right]^m 
\left[\begin{array}{c}
\ds\frac1{\sqrt N}\\ 
\noalign{\smallskip}
\sqrt{\ds\frac{N-1}N}\end{array}\right]\qquad \left(\theta 
\equiv \sin^{-1} \frac{2\sqrt{N-1}}{N}\right)\\
&= [1~~~0] \left[\begin{array}{cr}
\cos m\theta&\sin  m\theta\\ 
 -\sin m\theta&\cos m\theta\end{array}\right] 
\left[\begin{array}{c}
\ds\frac1{\sqrt N}\\ 
\noalign{\smallskip}
\sqrt{\ds\frac{N-1}N}\end{array}\right]\\
&= \frac1{\sqrt N} \cos m \theta + 
  \sqrt{\frac{N-1}N} \sin m\theta,
\end{align*}
or
\begin{equation}\label{grov:eq2.8'}
\langle w|U^m|s\rangle
= \cos (m\theta-\alpha),
 \qquad \alpha \equiv \cos^{-1} \frac1{\sqrt N}\,.
\end{equation}
Setting $\cos^2 (m\theta - \alpha) = 1$, 
  we can maximize the amplitude of 
$U^m|s\rangle$ in $|w\rangle$; thus 
\begin{align}
m\theta - \alpha &= 0,\nonumber\\ 
\label{grov:eq2.9}
m &= \frac{\alpha}\theta\,.    
\end{align}
 (If no integer satisfies this equation exactly, take the closest 
one.)
When $N$ is large, $\theta \approx \frac2{\sqrt N}$,
  $\alpha \approx \frac\pi2$, 
from (\ref{grov:eq2.9}) we obtain
\begin{equation}\label{grov:eq2.10}
m \approx \frac\pi2\Big/\left(\frac2{\sqrt N}\right) = \frac\pi4 \sqrt N.
\end{equation}
Therefore, after $m=\mathcal{O}(\sqrt{N})$ trials the state $|\,w\rangle$ 
will be projected out, which is precisely Grover's result. 
 By observing the qubits, we will 
 learn~$w$.
{\em By constructive interference, we have constructed 
$|\,w\rangle$}!
(Since $m$ only approximately satisfies (\ref{grov:eq2.9}),
 there is a small chance of getting a ``bad'' $w$.
 But because evaluating $f(w)$ is easy, in that case one will 
recognize the mistake and start over.)

\section{ Generalization of Grover's 
Algorithm to Multiobject Search}\label{grov:sec3}

\setcounter{equation}{0}

\indent

Here we generalize Grover's search algorithm in its original 
form [\ref{grover1}, 
\ref{grover2}] to the situation where the number of objects 
satisfying the search criterion is greater than 1.

Let a database $\{w_i\mid i=1,2,\ldots, N\}$, with corresponding orthonormal 
eigenstates 
$\{|w_i\rangle \,:\, i=1,2,\ldots, N\}$ in the QC, be given. 
 Let $f$ be an oracle 
function such that
$$f(w_j) = \left\{\begin{array}{ll}
1,&j=1,2,\ldots,\ell,\\
\noalign{\smallskip}
0&j=\ell+1, \ell+2,\ldots, N.\end{array}\right.$$
Here the $\ell$ elements $\{w_j\mid 1\le j\le \ell\}$ 
 are the desired objects of search.
 (To avoid introducing another layer of subscripts, we pretend in 
this theoretical discussion that these good objects are the first 
$\ell$ items in the list.
 In a real search application they would appear in the list in 
random order;
 in other words, all $N$ items $w_i$ are subjected to some unknown 
permutation, which we do not indicate explicitly.)
  Let
$\mathcal{H}$ be  the Hilbert  space generated by the orthonormal basis
$\mathcal{B} =  \{|w_j\rangle\mid j=1,\ldots, N\}$.
 Let $L =
\text{span}\{|w_j\rangle \mid  1\le j \le \ell\}$ be the subspace of
$\mathcal{H}$ spanned by the vectors of the good objects.

Define a linear operation in terms of the oracle function $f$ as follows:
\begin{equation}\label{grov:eq3.1}
I_L|w_j\rangle = (-1)^{f(w_j)} |w_j\rangle,\qquad j=1,2,\ldots, N.
\end{equation}
Then since $I_L$ is linear, the extension of $I_L$ to the entire space 
$\mathcal{H}$ is unique, with an ``explicit'' representation
\begin{equation}\label{grov:eq3.2}
I_L = \pmb{I} - 2 \sum^\ell_{j=1} |w_j\rangle \langle w_j|,
\end{equation}
where $\pmb{I}$ is the identity operator on $\mathcal{H}$. $I_L$ is 
the operator of {\em rotation (by $\pi$) of the phase\/} of the 
subspace $L$. Note again that the explicitness of 
(\ref{grov:eq3.2}) is misleading because explicit knowledge of 
$\{|w_j\rangle\mid 1\le j \le \ell\}$  in 
(\ref{grov:eq3.2}) is not available. Nevertheless, 
(\ref{grov:eq3.2}) is a well-defined (and  unitary) operator on 
$\mathcal{H}$ because of (\ref{grov:eq3.1}). (Unitarity
is a requirement for all operations in a~QC.) 

We now again define $|s\rangle$ as 
\begin{equation}\label{grov:eq3.3}
|s\rangle = \frac1{\sqrt N} \sum^N_{i=1} |w_i\rangle = \frac1{\sqrt N} 
\sum^\ell_{i=1} |w_i\rangle + \sqrt{\frac{N-\ell}N} |r\rangle,
\end{equation}
where now
$$|r\rangle = \frac1{\sqrt{1-(\ell/N)}} \left(|s\rangle - \frac1{\sqrt N}
\sum^\ell_{i=1} |w_i\rangle\right).$$
As before, we use 
\begin{equation}\label{grov:eq3.4}
I_s = \pmb{I} -2|s\rangle \langle s|.
\end{equation}
Note that $I_s$ in (\ref{grov:eq3.4}) is unitary and 
 hence quantum-mechanically 
admissible.  $I_s$ is  explicitly known,
 constructible with the so-called Walsh--Hadamard transformation.

\begin{lem}\label{grov:lem3.1}
Let $\wt L= \text{\rm span}(L\cup\{|r\rangle\})$.  Then $\{|w_i\rangle, 
|r\rangle\mid 
i=1,2,\ldots, \ell\}$ forms an orthonormal basis of $\wt L$. The orthogonal 
direct sum $\mathcal{H} = \wt L \oplus \wt L^\bot$ is an orthogonal invariant 
decomposition for both operators $I_{\wt L}$ and $I_s$. Furthermore,
\begin{itemize}
\item[(i)] The restriction of $I_s$ to $\wt L$ admits this 
 real unitary matrix 
representation with respect to the orthonormal basis $\{|w_1\rangle, 
|w_2\rangle,\ldots, |w_\ell\rangle, |r\rangle\}$:
\begin{align}
A &= [a_{ij}]_{(\ell+1)\times (\ell+1)},\nonumber\\
 \noalign{\smallskip}
\label{grov:eq3.5}
a_{ij} &= \left\{\begin{array}{ll}
\delta_{ij} - \ds\frac2N\,,&1\le i,j\le \ell,\\
 \noalign{\smallskip}
-\ds\frac{2\sqrt{N-\ell}}N (\delta_{i,\ell+1} + \delta_{j,\ell+1}),& i=\ell+1 
\text{ \rm or } j=\ell+1, i\ne j,\\
 \noalign{\smallskip}
\ds\frac{2\ell}N -1,&i=j=\ell+1.\end{array}\right.
\end{align}
\item[(ii)] The restriction of $I_s$ of $\wt L^\bot$ is $\mathbb{P}_{\wt 
L^\bot}$, 
the orthogonal projection operator onto $\wt L^\bot$. 
 Consequently, $I_s|_{\wt 
L^\bot} = \pmb{I}_{\wt L^\bot}$, where $\pmb{I}_{\wt L^\bot}$ is the identity 
operator on $\wt L^\bot$.
\end{itemize}
\end{lem}

\begin{proof}
We have, from (\ref{grov:eq3.3}) and (\ref{grov:eq3.4}),
\begin{align}
I_s &= \pmb{I} -2\left[\frac1{\sqrt N} \sum^\ell_{i=1} |w_i\rangle + 
\sqrt{\frac{N-\ell}N} |r\rangle\right] \left[\frac1{\sqrt N} \sum^\ell_{j=1} 
\langle w_j| + \sqrt{\frac{N-\ell}N} \langle r|\right]\nonumber\\
&= \left[\sum^\ell_{i=1} |w_i\rangle \langle w_i\rangle + |r\rangle \langle r| + 
\mathbb{P}_{\wt L^\bot}\right] - \left\{\frac2N \sum^\ell_{i=1} \sum^\ell_{j=1} 
|w_i\rangle \langle w_j|\right.\nonumber\\
&\quad \left. + \frac{2\sqrt{N-\ell}}N \left[\sum^\ell_{i=1} (|w_i\rangle 
\langle r| + |r\rangle \langle w_i|)\right] + 2\left(\frac{N-\ell}N\right) 
|r\rangle \langle r|\right\}\nonumber\\
&= \sum^\ell_{i=1} \sum^\ell_{j=1} \left(\delta_{ij} - \frac2N\right) 
|w_i\rangle \langle w_j| - \frac{2\sqrt{N-\ell}}N \left[\sum^\ell_{i=1} 
(|w_i\rangle \langle r| + |r\rangle \langle w_i|)\right]\nonumber\\
\label{grov:eq3.6}
&\quad + \left(\frac{2\ell}N-1\right) |r\rangle \langle r| + \mathbb{P}_{\wt 
L^\bot}.
\end{align}
The conclusion follows.
\end{proof}

The generalized ``Grover search engine'' for multiobject search is now 
 constructed as
\begin{equation}\label{grov:eq3.7}
U  = -I_sI_L\,.
\end{equation}

\begin{lem}\label{grov:lem3.2}
The orthogonal direct sum $\mathcal{H} = \wt L \oplus \wt L^\bot$ is an 
invariant decomposition for the unitary operator $U$, such that the
following 
holds:
\begin{itemize}
\item[(1)] With respect to the orthonormal basis $\{|w_1\rangle,\ldots, 
|w_\ell\rangle, |r\rangle\}$ of $\wt L$, $U$ 
 admits the real unitary matrix representation
\begin{align}
U|_{\wt L} &= [u_{ij}]_{(\ell+1)\times (\ell+1)},\nonumber\\
 \noalign{\smallskip}
\label{grov:eq3.8}
u_{ij} &= \left\{\begin{array}{ll}
\delta_{ij} - \ds\frac2N\,,&1\le i,j\le \ell,\\
  \noalign{\smallskip}
\ds\frac{2\sqrt{N-\ell}}N (\delta_{j,\ell+1} - \delta_{i,\ell+1}), 
 &i=\ell+1 
\text{ \rm or } j=\ell+1, i\ne j,\\
 \noalign{\smallskip}
1-\ds\frac{2\ell}N\,,&i=j=\ell+1.\end{array}\right.  
\end{align}
\item[(2)] The restriction of $U$ to $\wt L^\bot$ is $-\mathbb{P}_{\wt 
L^\bot} = -\pmb{I}_{\wt L^\bot}\,$.
\end{itemize}
\end{lem}

 \goodbreak
\begin{proof}
Substituting (\ref{grov:eq3.2}) and (\ref{grov:eq3.6}) into (\ref{grov:eq3.7}) 
and simplifying, we obtain
\begin{align*}
U &= -I_sI_L = \cdots \text{(simplification)}\\
&= \sum^\ell_{i=1} \sum^\ell_{j=1} \left(\delta_{ij} - \frac2N\right) 
|w_i\rangle \langle w_j| + \frac{2\sqrt{N-\ell}}N \sum^\ell_{i=1} (|w_i\rangle 
\langle r| - |r\rangle \langle w_i|)\\
&\quad + \left(1 - \frac{2\ell}N\right) |r\rangle \langle r| - \mathbb{P}_{\wt 
L^\bot}\,.
\end{align*}
The lemma follows.
\end{proof}

Lemmas \ref{grov:lem3.1} and \ref{grov:lem3.2} above effect a 
reduction of the problem to an invariant subspace $\wt L$. However, 
$\wt L$ is an $(\ell+1)$-dimensional subspace where $\ell$ may also 
be fairly large. 
 Another reduction of dimensionality is needed to further simplify the 
operator $U$.

\begin{prop}\label{grov:prop3.3}
Define $\mathcal{V}$ by
$$\mathcal{V} = \left\{|v\rangle \in \wt L\,:\,
  |v\rangle = a \sum^\ell_{i=1} 
|w_i\rangle + b|r\rangle;\,\, a, b\in \mathbb{C}\right\}.$$
 Then $\mathcal{V}$ is an 
invariant two-dimensional subspace of $U$ such that
\begin{itemize}
\item[(1)] $r,s\in \mathcal{V}$;
\item[(2)] $U(\mathcal{V}) = \mathcal{V}$.
\end{itemize}
\end{prop}

\begin{proof}
Straightforward verification.
\end{proof}

Let $|\wt w\rangle = \frac1{\sqrt\ell} \sum\limits^\ell_{i=1} 
|w_i\rangle$. Then $\{|\wt w\rangle, |r\rangle\}$ forms an 
orthonormal basis of $\mathcal{V}$. We have the second reduction, 
to dimensionality 2. 

\begin{thm}\label{grov:thm3.4} 
With respect to the orthonormal basis $\{|\wt w\rangle, 
|r\rangle\}$ in the invariant subspace $\mathcal{V}$, $U$ admits 
the real unitary matrix representation 
\begin{equation}\label{grov:eq3.9} U =
  \left[\begin{matrix} \frac{N-
2\ell}N& \frac{2\sqrt{\ell (N-\ell)}}N\\ 
\noalign{\smallskip}
 \frac{-2\sqrt{\ell(N-
\ell)}}N& \frac{N-2\ell}N\end{matrix}\right] = 
\left[\begin{matrix} \cos \theta & \sin \theta\\
 - \sin \theta& \cos\theta 
\end{matrix}\right],\quad \theta\equiv \sin ^{-1} 
 \left(\frac{2\sqrt{\ell(N-\ell)}}N 
\right).
\end{equation}
\end{thm}

\begin{proof}
Use the matrix representation (\ref{grov:eq3.8}) and the definition of
$|\wt w\rangle$.
\end{proof}

Since $|s\rangle\in \mathcal{V}$, we can calculate $U^m|s\rangle$ efficiently 
using (\ref{grov:eq3.9}):
\begin{align}
U^m|s\rangle &= U^m \left(\frac1{\sqrt N} \sum^\ell_{i=1} 
|w_i\rangle + 
\sqrt{\frac{N-\ell}N} |r\rangle\right)\qquad \text{(by (\ref{grov:eq3.3}))} 
\nonumber\\
&= U^m \left(\sqrt{\frac\ell N} 
|\wt w\rangle + \sqrt{\frac{N-\ell}N} 
|r\rangle\right)\nonumber\\
&= \left[\begin{matrix} \cos \theta& \sin \theta\\
 - \sin \theta&\cos 
\theta\end{matrix}\right]^m \left[\begin{matrix} \sqrt{\frac\ell N}\\ 
\noalign{\smallskip}
\sqrt{\frac{N-\ell}N}\end{matrix}\right]\nonumber\\
\label{grov:eq3.10}
&= \left[\begin{matrix} \cos (m\theta-\alpha)\\ 
- \sin(m\theta -\alpha)] 
\end{matrix}\right]\quad \left(\alpha \equiv \cos^{-1} 
\sqrt{\frac\ell N}\right),\\
&= \cos(m \theta-\alpha)\cdot |\wt w\rangle 
 - \sin(m \theta - \alpha)\cdot 
|r\rangle.\nonumber
\end{align}
Thus, the probability of reaching the state $|\wt w\rangle$ 
 after $m$ iterations 
is
\begin{equation}\label{grov:eq3.11}
P_m = \cos^2(m\theta -\alpha). 
\end{equation}
 If $\ell \ll N$, then 
$\alpha$ is close to $\pi/2$ and, therefore, (\ref{grov:eq3.11}) 
is an 
increasing function of $m$ initially. This again manifests the notion of 
amplitude amplification. This probability $P_m$ is maximized if 
 $m\theta -\alpha 
= 0$, implying 
$$m = \left[\frac{\alpha}\theta\right] = \text{the integral part of } 
\frac{\alpha}\theta\,.$$ 
When $\ell/N$ is small, we have
\begin{align*}
\theta &= \sin^{-1} \left(\frac{2\sqrt{\ell(N-\ell)}}N\right)\\
&= \sin^{-1} \left(2 \sqrt{\frac\ell{N}} \left[1-\frac12 \frac\ell{N} - \frac18 
\left(\frac\ell{N}\right)^2 \pm\cdots\right]\right)\\
&= 2 \sqrt{\frac\ell{N}} + \mathcal{O}((\ell/N)^{3/2}),\\
\alpha &= \cos^{-1} \sqrt{\frac\ell N} = \frac\pi2 - \left[\sqrt{\frac\ell N} + 
\mathcal{O} ((\ell/N)^{3/2})\right].
\end{align*}
Therefore
\begin{align}
m &\approx \frac{\frac\pi2 - \left[\sqrt{\frac\ell N} + 
\mathcal{O}((\ell/N)^{3/2})\right]} 
 {2\sqrt{\frac\ell{N}} + \mathcal{O} 
((\ell/N)^{3/2})}\nonumber\\
\label{grov:eq3.12}
&= \frac\pi4 \sqrt{\frac{N}\ell} \left[1 + \mathcal{O}\left(\frac\ell{N} 
\right)\right].
\end{align}

\begin{cor}\label{grov:cor3.5}
The generalized Grover algorithm for  multiobject search with operator 
$U$ 
given by (\ref{grov:eq3.7}) has success probability $P_m = \cos^2(m\theta 
-\alpha)$ 
 of reaching the state $|\wt w\rangle \in L$ after $m$ 
iterations. For $\ell/N$ small, after $m = \frac\pi4 \sqrt{N/\ell}$ iterations, 
the probability of reaching $|\wt w\rangle$ is close to 1.$\hfill\square$
\end{cor}

The result (\ref{grov:eq3.12}) is consistent with Grover's original algorithm 
for single object search with $\ell=1$, which has $m\approx \frac\pi4 \sqrt N$; 
cf.\ (\ref{grov:eq2.10}).

\begin{thm}\label{grov:thm3.6} {\bf (Boyer, Brassard, H\o yer and Tapp 
[\ref{BBHT}]).}
Assume that $\ell/N$ is small. Then any search algorithm for $\ell$ objects, in 
the form of
$$U_pU_{p-1}\ldots U_1|w_I\rangle,$$
where each $U_j$, $j=1,2,\ldots, p$, is a unitary operator and $|w_I\rangle$ is an 
arbitrary superposition state, takes in average $p=\mathcal{O}(\sqrt{N/\ell})$ 
iterations in order to reach the subspace $L$ with a positive probability 
$P>\frac12$ 
independent of $N$ and $\ell$. 
Therefore, the generalized Grover algorithm in 
 Corollary~\ref{grov:cor3.5} is of 
optimal order.
\end{thm}

\begin{proof}
This is the major theorem in [\ref{BBHT}]; see Section~7 and particularly
Theorem~8 therein.
 Note also the work of Zalka~[\ref{zalka}].
\end{proof}

Unfortunately, if the number $\ell$ of good items is not known in 
advance, Corollary~\ref{grov:cor3.5} does not tell us when to stop 
the iteration.
 This problem was addressed in~[\ref{BBHT}], and in another way 
 in~[\ref{BHT}].
 In a related context an equation arose that was not 
fully solved in~[\ref{BBHT}].
We consider it in the final segment of this paper.
   As in [\ref{BBHT},~\S 3], consider stopping the Grover process 
   after $j$ iterations, and, if a good object is not obtained, 
 starting it over again from the beginning.
  From Corollary~\ref{grov:cor3.5}, the probability of success 
  after $j$ iterations is $\cos^2(j\theta -\alpha)$.
 By a well-known theorem of probability theory, if the probability 
of success in one ``trial'' is~$p$,
 then the expected number of trials before success is achieved 
will be $1/p$.
(The probability that success is achieved on the $k$th trial
is $p(1-p)^{k-1}$.
Therefore, the expected number of trials is
\begin{equation}
\sum_{k=1}^\infty kp(1-p)^{k-1} =
-p \sum_{k=1}^\infty \frac d{dp} (1-p)^k =
-p\, \frac d{dp} \frac{1-p}p \,,
\end{equation}
which is $1/p$.)
 In our case, each trial consists of $j$ Grover iterations, so the 
expected number of iterations before success is
$$E(j) = j\cdot \sec^2 (j\theta-\alpha).$$  
The optimal number of iterations $j$ is obtained by
 setting the derivative $E'(j)$ equal to zero: 
\begin{align}
0 = E'(j) &= \sec^2(j\theta-\alpha)
  +2j\theta\sec^2(j\theta-\alpha)  
\tan(j\theta-\alpha),\nonumber\\ 
\label{grov:eq3.13}
2j\theta &= -\cot((j\theta-\alpha)). 
\end{align}
 (In [\ref{BBHT}, \S 3], 
 this 
equation is derived 
 in the form $4\vartheta j = \tan((2j+1)\vartheta)$, 
 which is seen to be equivalent to (\ref{grov:eq3.13}) by noting 
 that $\vartheta = \frac{\theta}2 = \frac{\pi}2 - \alpha$. 
Those authors then note that they have not solved the equation 
 $4\vartheta j = \tan((2j+1)\vartheta)$ 
 but proceed to use an ad hoc equation $z = \tan(z/2)$ with 
$z=4\vartheta j$ instead.) 
 Let us now approximate the solution $j$ of 
(\ref{grov:eq3.13}) 
iteratively as follows. From (\ref{grov:eq3.13}),
\begin{align}
&2j\theta \sin(j\theta-\alpha) + \cos(j\theta-\alpha)
  = 0,\nonumber\\
\label{grov:eq3.14}
&e^{2i(\theta j-\alpha)} = (i2\theta j+1)/(i2\theta j-1), 
\end{align}
and by taking the logarithm of both sides, we obtain
\begin{equation}\label{grov:eq3.15}
2i(\theta j-\alpha) = 2i\pi n 
 + i\arg\left(\frac{i2\theta j+1}{i2\theta 
j-1}\right) + \ln\left|\frac{i2\theta j+1}{i2\theta j-1}\right|,
\end{equation}
for any integer $n$. Assume that $\ell/N$ is small so that $j$ is large, but we 
are looking for the smallest such positive $j$. Note that the logarithmic term 
in (\ref{grov:eq3.15}) vanishes, and
\begin{align*}
\arg\left(\frac{i2\theta j+1}{i2\theta j-1}\right) &=
 - 2\tan^{-1} \frac1{2\theta 
j}\\
&= 2\left[\sum^\infty_{q=0} \frac{(-1)^{q+1}}{2q+1}
  \left(\frac1{2\theta 
j}\right)^{2q+1}\right]\\
&=-\, \frac1{\theta j} + \mathcal{O}((\theta j)^{-3}); 
\end{align*}
by taking $n=0$ in (\ref{grov:eq3.15}), we obtain
\begin{align}
j &= \frac1{2i\theta} \left[2i\alpha 
  - i\cdot \frac1{\theta j} + 
\mathcal{O} ((\theta j)^{-3})\right]\nonumber\\
\label{grov:eq3.16}
&= \frac1\theta \left[\alpha  - \frac1{2\theta j} + 
\mathcal{O}((\theta j)^{-3})\right].
\end{align}

The first order approximation $j_1$ for $j$ is obtained by solving
\begin{align}
j_1 &= \frac1\theta \left(\alpha
  - \frac1{2\theta j_1}\right), 
\nonumber\\
j^2_1 &- \frac1\theta\,  \alpha j_1
 + \frac1{2\theta^2} =0, 
\nonumber\\
\label{grov:eq3.17}
j_1 &= \frac1{2\theta} (\alpha + 
\sqrt{\alpha^2-2}). 
\end{align}
Higher order approximations $j_{n+1}$ for $n=1,2,\ldots,$ may be obtained by 
successive iterations
$$j_{n+1} = \frac1\theta \left(\alpha  - \tan^{-1} 
\frac1{2\theta j_n}\right)$$
based on (\ref{grov:eq3.13}). This process will yield a convergent 
solution $j$ to (\ref{grov:eq3.13}).

\medskip
{\bf Acknowledgments:} 
We thank B.-G. Englert and M. Hillery for acquainting us with
some of the literature of quantum computation,  M.~M.~Kash
for a technical discussion, and Hwang~Lee, J.~D.~Malley, and
D.~A.~Lidar for comments on the manuscript.

\end{document}